\documentstyle[prl,aps]{revtex}

\begin{document}
\author{Jian Qi Shen $^{1,}$$^{2}$ \footnote{E-mail address: jqshen@coer.zju.edu.cn}}
\address{$^{1}$  Centre for Optical
and Electromagnetic Research, Joint Research Centre of Photonics
of The Royal Institute of Technology (Sweden) and Zhejiang
University, Zhejiang University, Hangzhou Yuquan 310027, P.R. China\\
$^{2}$ Zhejiang Institute of Modern Physics and Department of
Physics, Zhejiang University, Hangzhou 310027, P.R. China}
\date{\today }
\title{The evolution of atomic matter waves inside an atom fiber}
\maketitle

\begin{abstract}
A phenomenological description of time evolution of atomic matter
waves inside a spiral shaped atomic-wave guide is presented in
this report. We study three related topics: (i) the effective
Hamiltonian and the time-development equation governing the matter
waves; (ii) wavefunctions describing the matter waves in the
noncoplanar atom guide; (iii) showing that such wavefunctions
obtained is just the eigenstates of the atomic momentum operator.
It is believed that both the idea and the results presented here
may have relevance to some fundamental problems of {\it atom
optics}.

PACS number(s): 03.65.Ca, 32.80.Pj
\end{abstract}
\pacs{03.65.Ca, 32.80.Pj}

Recently, the implementation of atom chips based on the
microfabrication techniques\cite{Folman,Reichel,Kruger} in atom
optics can allow trapping and guiding of atoms with a high
accuracy, and therefore has many
applications\cite{Reichel2,Folman2,Cassettari} in mesoscopic
physics. More recently, Luo {\it et al.} reported a novel work of
the so-called ``atom fiber'', {\it i.e.}, an atomic guide that can
control the matter waves (such as cold neutral atoms and ideally
Bose-Einstein condensates) on a two dimensional surface of the
atom chip\cite{Luo}. In their experiments, Luo {\it et al.}
demonstrated guiding atoms around more than two complete turns
along a spiral shaped 25 mm long curved path at various
atom-surface distances\cite{Luo}. Such a technique is a key
element for the controlled manipulation of matter waves on the
atom chip. It is believed that the atom fiber and related
technologies may have some potential applications to, {\it e.g.},
the quantum registers for quantum information processing and the
guided matter wave interferometers ({\it e.g.}, the wide angle
Mach-Zehnder and Sagnac interferometers)\cite{Luo}.

In Luo {\it et al.}'s paper, investigators presented the
experimental details and mechanism for the realization of the
atomic matter-wave guide\cite{Luo}. Because of the complexity of
the conditions for realizing the atom fiber, the theoretical
description of wave propagation of atoms inside it can be achieved
neither in a straightforward way nor from the first principle. In
this report we demonstrate that based on some assumptions, which
can simplify the interaction details of atomic matter wave on the
atom chip, there may exist a phenomenological description that
will give an enlightening understanding of the time evolution of
matter waves inside the curved atomic wave guide.

In what follows, by ignoring the subordinate details of state
evolution in an atom fiber, we assume that the atom fiber can be
viewed as a perfect guide, namely, (i) the guide does not alter
the magnitude but the direction of atom wave vector ${\bf k}$ in
the evolution process, and (ii) from the point of view of
mesoscopic scale, the wave vector ${\bf k}$ of atomic matter waves
may be said to be always along the tangent to the curved guide
(path) at each point at arbitrary time in the evolution process
inside the atom fiber. Note that here the second assumption can
truly hold in this process since, for example, in the side guide
that is the simplest form of a magnetic guide, atoms can only move
along the third unconfined direction as long as the external bias
field points in the direction perpendicular to the guiding
direction\cite{Luo}. In this sense, the wave vector of atoms can
characterize the geometric shape of the wire guide, which enables
us to deal with the propagation problem of atomic matter waves in
the spiral shaped wire guide. The aim of this report is to develop
a phenomenological method for treating the time evolution of atom
wavefunctions based on these two assumptions in such atom fibers.

According to the above two assumptions, the momentum (wave vector)
squared ${\bf k}^2$ of a matter wave propagating inside a
noncoplanarly curved atom fiber is conserved, which will yield the
following identity
\begin{equation}
\dot{\bf{k}}+{\bf{k}}\times \left(\frac{{\bf{k}}\times
\dot{\bf{k}}}{k^{2}}\right)=0,     \label{1}
\end{equation}
where dot denotes the derivative with respect to time, and $k$ is
the magnitude of ${\bf k}$, {\it i.e.}, $k=|{\bf k}|$. Although
Luo {\it et al.}'s atom fiber was the coplanar shaped wire guide
on the chip surface ({\it i.e.}, the spiral guild has a
two-dimensional wire plane), instead, here we will consider a more
general case in which the atom fiber is noncoplanarly shaped.
This, therefore, means that we will not restrict the atom wave
vector only to the two-dimensional spiral shaped fiber ({\it
i.e.}, in the following discussion ${\bf k}$ can be a three
dimensional vector). Since Eq.(\ref{1}) is exactly analogous to
the equation of motion of a charged particle moving in a magnetic
field or of a spinning particle in a rotating frame of reference,
${{\bf{k}}\times \dot{\bf{k}}}/{k^{2}}$ can be regarded as a
``magnetic field'' or ``gravitomagnetic field''\cite{Shen,Shenprb}
(correspondingly, ${\bf{k}}\times ({{\bf{k}}\times
\dot{\bf{k}}})/{k^{2}}$ can be thought of as a ``Lorentz magnetic
force'' or ``Coriolis force''). Just similar to Mashhoon's work of
deriving the interaction Hamiltonian of gravitomagnetic dipole
moment ({\it i.e.}, spin) in a gravitomagnetic field
\cite{Mashhoon3,Mashhoon1}, one can also readily obtain the
Hamiltonian describing the coupling of the atomic total
``gravitomagnetic moment'' ({\it i.e.}, the total angular momentum
operator ${\bf J}$) to the ``gravitomagnetic field'' as follows
\begin{equation}
H(t)=\frac{{\bf{k}}(t)\times \dot{\bf{k}}(t)}{k^{2}}\cdot
{\bf{J}}, \label{2}
\end{equation}
which may be the effective Hamiltonian describing the propagation
of atomic matter waves in the atom fiber.

The above derivation of the effective Hamiltonian is enlightening,
the physical meanings of which is explicit. Now we will treat this
problem in a more rigorous way, {\it i.e.}, by using the
infinitesimal rotation operator of atom wavefunction. The atom
wavefunction $|m, {\bf k}(t)\rangle$ (with $m$ being the
eigenvalue of the third component of ${\bf J}$) varies as it
rotates by an infinitesimal angle, say $\vec{\vartheta}$, namely,
it obeys the following transformation rule (in the unit $\hbar=1$)
\begin{equation}
|m, {\bf k'}(t)\rangle=\exp\left[-i\vec{\vartheta}\cdot{\bf
J}\right]|m, {\bf k}(t)\rangle,
\label{eqA1}
\end{equation}
where $\exp\left[-i\vec{\vartheta}\cdot{\bf J}\right]\simeq
1-i\vec{\vartheta}\cdot{\bf J}$, and ${\bf k'}(t)={\bf
k}(t)+\Delta {\bf k}(t)$ with $\Delta {\bf
k}(t)=\dot{\bf{k}}{\Delta t}$. Here $|\vec{\vartheta}|$ is the
angle between ${\bf k}(t)$ and ${\bf k'}(t)$, and the direction of
$\vec{\vartheta}$ is parallel to that of ${\bf k}(t)\times{\bf
k'}(t)$. One can therefore arrive at
\begin{equation}
\vec{\vartheta}=\frac{{\bf k}(t)\times{\bf k'}(t)}{k^2}=\frac{{\bf
k}(t)\times\dot{\bf{k}}(t)}{k^2}{\Delta t}.
 \label{eqApp2}
\end{equation}
Expanding $|m, {\bf k'}(t)\rangle$ in (\ref{eqA1}) to the first
order, it follows that
\begin{equation}
i\frac{\partial \left|m,{\bf{k}}(t)\right\rangle }{\partial
t}=\frac{{\bf{k}}(t)\times \dot{\bf{k}}(t)}{k^{2}}\cdot
{\bf{J}}\left|m,{\bf{k}}(t)\right\rangle.        \label{eqA3}
\end{equation}
Not that Eq.(\ref{eqA3}) is just the form of the time-dependent
Schr\"{o}dinger equation, which governs the evolution of atomic
matter waves inside a noncoplanar atom fiber. Thus by using the
technique of infinitesimal rotation of atom wavefunctions, we
obtain the effective Hamiltonian (\ref{2}) (and hence the
time-dependent Schr\"{o}dinger equation) of atoms moving inside
the curved atomic-wave guild.

It follows from Eq.(\ref{eqA3}) that the atom fiber with atomic
matter waves inside it is just the three-generator Hamiltonian
system, the interaction Hamiltonian of which is the linear
combination of three algebraic generators which form a certain Lie
algebra, say, $su(2)$ for the present system. In the literature,
we have obtained the exact solutions of the arbitrary
three-generator time-dependent quantum systems (and
models)\cite{Shen2}. Based on this work\cite{Shen2}, here we will
straightforwardly write the wavefunctions of Eq.(\ref{eqA3})
without presenting the details of derivation procedure. For
convenience, in what follows the atom wave vector is rewritten in
the spherical polar coordinate system, {\it i.e.}, ${\bf
k}(t)=k(\sin\theta\cos\varphi, \sin\theta\sin\varphi,
\cos\theta)$, where both $\theta$ and $\varphi$ are the
time-dependent functions. We assume that the initial wave vector
is ${\bf k}(0)=(0, 0, k)$, {\it i.e.}, the initial polar angle
$\theta(0)=0$. According to the results obtained in the
reference\cite{Shen2}, the wavefunction in Eq.(\ref{eqA3}) can be
written in the form

\begin{equation}
|m, {\bf
k}(t)\rangle=\exp\left[\frac{1}{i}\phi_{m}(t)\right]V(t)|m\rangle,
\label{wavefunction}
\end{equation}
where $|m\rangle$ is the initial state satisfying the eigenvalue
equation $J_{3}|m\rangle=m|m\rangle$, and the expression for the
phase is
$\phi_{m}(t)=m\int^{t}_{0}\dot{\varphi}(t')\left[1-\cos\theta
(t')\right]{\rm d}t'$. The time-dependent operator $V(t)$ takes
the form
\begin{equation}
V(t)=\exp\left[\beta(t) J_{+}-\beta^{\ast}(t) J_{-}\right]
\end{equation}
with $\beta=-(\theta/2)\exp(-i\varphi)$ and
$\beta^{\ast}=-(\theta/2)\exp(i\varphi)$. Here the operators
$J_{\pm}=J_{1}\pm iJ_{2}$.

It is apparently seen that the momentum operator ${\bf p}$ of
atoms is also a physical quantity that can characterize the
propagation of atomic matter waves in the coiled atom fiber, since
in accordance with the second assumption, the direction of wave
vector ${\bf k}(t)$ (the time-dependent eigenvalue of ${\bf p}$)
contains the information on the geometric shape of the curved wire
guide. In order to show that our treatment of evolution of atomic
matter waves in the atom fiber is self-consistent, we should prove
that the above wavefunction is the eigenstate of the momentum
operator of atoms in the evolution process. In the Cartesian
coordinate system, by the aid of the following commuting relations
\begin{equation}
[J_{i}, p_{j}]=i\epsilon_{ijk}p_{k}
\end{equation}
with $\epsilon_{ijk}$ being the three dimensional Levi-Civita
tensor, one can arrive at
\begin{eqnarray}
& &  V^{\dagger}(t)p_{1}V(t)=p_{1}+\sin\theta\cos\varphi p_{3}+\frac{\left(1-\cos\theta\right)\cos\varphi}{\theta}\left(\beta p_{+}+\beta^{\ast}p_{-}\right),                  \nonumber \\
& &  V^{\dagger}(t)p_{2}V(t)=p_{2}+\sin\theta\sin\varphi p_{3}+\frac{\left(1-\cos\theta\right)\sin\varphi}{\theta}\left(\beta p_{+}+\beta^{\ast}p_{-}\right),                     \nonumber \\
& &
V^{\dagger}(t)p_{3}V(t)=p_{3}+\left(\cos\theta-1\right)p_{3}+\frac{\sin\theta}{\theta}\left(\beta
p_{+}+\beta^{\ast}p_{-}\right),
\end{eqnarray}
where $p_{\pm}=p_{1}\pm ip_{2}$. With the help of these relations,
one can obtain
\begin{equation}
p_{i}|m, {\bf
k}(t)\rangle=\left[k\sin\theta(\delta_{1i}\cos\varphi+\delta_{2i}\sin\varphi)-k\left(1-\cos\theta\right)\delta_{3i}+k_{i}\right]|m,
{\bf k}(t)\rangle.     \label{eigen}
\end{equation}
Here $k_{i}$ denotes the initial momentum at $t=0$. That is,
$k_{1}(0)=0$, $k_{2}(0)=0$ and $k_{3}(0)=k\cos\theta(0)=k$. Thus
it follows from Eq.(\ref{eigen}) that
\begin{equation}
{\bf p}|m, {\bf k}(t)\rangle={\bf k}(t)|m, {\bf k}(t)\rangle,
\end{equation}
which is shown to be the eigenvalue equation of the momentum
operator of atoms in the curved atom fiber. Thus, the solution
$|m, {\bf k}(t)\rangle$ of Schr\"{o}dinger equation (\ref{eqA3})
is truly the eigenstate of the momentum operator ${\bf p}$ of
atoms.

In addition, the operator $I(t)={\bf k}(t)\cdot{\bf J}/k$, which
is the total angular momentum operator, ${\bf J}$, of atoms
projected onto the atom momentum (wave vector) ${\bf k}(t)$, is a
physically interesting quantity, since it agrees with the
Liouville-Von Neumann equation ${\partial I(t)}/{\partial
t}-i[I(t),H(t)]=0$. This, therefore, means that the operator ${\bf
k}(t)\cdot{\bf J}/k$ can be considered a Lewis-Riesenfeld
invariant\cite{Lewis}, which possesses the conserved eigenvalues
in the evolution process. Thus, for the atomic matter waves in the
atom fiber, it may be referred to as the atomic {\it helicity
operator}.

To conclude, assuming that the atomic matter wave is propagating
inside a perfect atom guide, we construct an effective Hamiltonian
to describe the interaction between the atomic matter waves and
the spiral shaped guide, and present the exact solutions of the
time-dependent Schr\"{o}dinger equation that governs the evolution
of matter wave in the coiled atom fiber (guide), as realized
experimentally by Luo {\it et al.}\cite{Luo}. Since it is shown
that such wavefunctions is truly the eigenstates of momentum
operator of atoms, we think that, at least within the framework of
the two assumptions made above, our method is self-consistent. So,
it may be believed that the formulation presented here would
provide us with a useful insight into the evolution problem in the
atom fibers, and therefore deserves further consideration in the
investigation of atom optics.

\textbf{Acknowledgements}  This project was supported partially by
the National Natural Science Foundation of China under Project No.
$90101024$ and $60378037$. I thank Q. Li and D. Wang for their
valuable suggestions and comments on my manuscript.

\end{document}